\documentclass[twocolumn,prl,tightenlines,superscriptaddress]{revtex4-2}

\usepackage{hyperref}

\usepackage{amsmath}
\usepackage{amssymb,amsfonts,latexsym}
\usepackage{bm}
\usepackage[mathcal]{euscript}
\usepackage{graphicx}
\usepackage{epsfig}
\usepackage{color}
\usepackage{xfrac}
\usepackage{mathrsfs}

\begin{document}

\title{Effect of Persistent Noise on the XY Model and Two-Dimensional Crystals}

\author{Xia-qing Shi}
\affiliation{Center for Soft Condensed Matter Physics and Interdisciplinary Research, Soochow University, Suzhou 215006, China}

\author{Hugues Chat\'{e}}
\affiliation{Service de Physique de l'Etat Condens\'e, CEA, CNRS Universit\'e Paris-Saclay, CEA-Saclay, 91191 Gif-sur-Yvette, France}
\affiliation{Computational Science Research Center, Beijing 100094, China}

\date{\today}

\begin{abstract}
Two-dimensional (2D) crystals made of active particles were shown recently to be able to
experience extremely large spontaneous deformations without melting. 
The root of this phenomenon was argued to lie in the time-persistence of the orientation of the 
intrinsic axes of particles. Here, we pursue this idea and consider passive systems subjected to time-persistent external perturbations.
We first study a 2D XY model subjected to time-correlated noise and find that it can remain quasi-ordered in spite of correlations decaying much faster than allowed in equilibrium. We then study a simple model of a passive 2D crystal
immersed in a bath of active particles, and show that it can sustain large deformations without melting.
\end{abstract}

\maketitle

Studies of active matter constitute an imposing and still fast-growing body or work. 
Whereas active fluids take the lion's share, crystalline arrangements of active particles have been
observed and investigated, mostly in two dimensions (2D), uncovering a wealth of interesting properties
\cite{
bialke2012crystallization,
redner2013structure,
singh2016universal,
thutupalli2018flow-induced,
cugliandolo2017phase,
digregorio2018full,
digregorio20192D,
klamser2018thermodynamic,
pasupalak2020hexatic,
paliwal2020role,
loewe2020solid,
digregorio2022unified,
gregoire2003moving,
ferrante2013elasticity,
ferrante2013collective,
menzel2013unidirectional,
menzel2014active,
weber2014defect-mediated,
rana2019tuning,
huang2021alignment,
briand2018spontaneously,
vanderlinden2019interrupted,
riedel2005self-organized,
sumino2012large,
nguyen2014emergent,
yeo2015collective,
goto2015purely,
petroff2015fast-moving,
oppenheimer2019rotating,
huang2020dynamical,
james2021emergence,
tan2022odd,
oppenheimer2022hyperuniformity,
vanzuiden2016spatiotemporal,
braverman2021topological}.
In such systems, the non-equilibrium character allowing for unusual properties is intrinsic to the particles
composing the crystal. 
Here we take a different approach and consider extrinsic sources of non-equilibriumness 
acting on a standard, `passive' crystal. 

In a recent publication \cite{shi2023extreme}, we showed that two-dimensional (2D) crystals made 
of active particles subjected to pairwise repulsion forces 
can experience extremely large spontaneous deformations without melting. 
Such active crystals exhibit long-range bond order and algebraically-decaying positional order,
but with an exponent $\eta$ not limited by the $\tfrac{1}{3}$ bound given by the (equilibrium) KTHNY theory
\cite{kosterlitz1972long,kosterlitz1973ordering,halperin1978theory,nelson1979dislocation,young1979melting,strandburg1988two-dimensional}.

These findings were rationalized using linear elastic theory and the existence of two well-defined 
effective temperatures quantifying respectively large-scale deformations and bond-order fluctuations.
For the simple case of a 2D crystal made of active Brownian particles moving at constant 
speed/self-propulsion force $s_0$ while their intrinsic axis is subjected to rotational diffusion 
of strength $D_r$,
large-scale elastic deformations are controlled by an effective temperature 
$T_S \sim s_0^2$. Fluctuations of (usually hexatic) bond-order, which are mostly small-scale, 
are also controlled by an effective temperature $T_6$ scaling like $s_0^2$.
In the cases studied in \cite{shi2023extreme}, adjusting $T_S$ and $T_6$ so that $T_S=T_6=T$ 
in the equilibrium $D_r\to\infty$ limit,
we showed that $T_6<T_S$ when going `away' from equilibrium. 
In such a situation, $T_S$ can reach large values before local fluctuations ---monitored via $T_6$--- are strong enough to create free defects that melt the lattice, allowing the exponent $\eta$ to take large values.

The root of these phenomena was argued to ultimately lie in the 
time-persistence of the orientation of intrinsic axes of the active particles forming the crystal. 
Here we go one step further and consider passive systems subjected to time-persistent external perturbations.
We first study the barest setting for such a situation:
a 2D XY model subjected to time-correlated noise. 
We show that the transition separating the region of quasi-long-range order from the disordered phase
is delayed by the time-correlations of the noise, allowing to observe much stronger spin waves than in equilibrium,
without proliferation of defects.
We then consider a direct consequence of the above ideas: a passive crystal immersed 
in a bath of active particles ---a case of possible experimental relevance--- and show that this also yields large deformations without melting, although with properties possibly qualitatively different from those of the simple
active crystals studied in \cite{shi2023extreme}.

{\it XY model with time-correlated noise.}
We consider a triangular lattice of spins $i$ whose orientation angles $\theta_i$ align locally and 
are subjected to time-correlated noise, which we choose to be of the Ornstein-Uhlenbeck type for simplicity.
The equation of motion for $\theta_i$ reads:
\begin{align}
&\dot{\theta}_i = \kappa  \langle\sin (\theta_j - \theta_i)\rangle_{j\sim i}  +\varpi_i \label{XY}\\
{\rm with} \; &\dot{\varpi_i} = -\frac{\varpi_i}{\tau} +\sqrt{{2T}/{\tau}} \,\xi_i \label{OU}
\end{align}
where $\kappa$ is the alignment elastic constant
\footnote{There are only two independent parameters, which we choose to be $T$ and $\tau$, fixing $\kappa=0.5$
without loss of generality.}
and neighbors $j$ of spin $i$ are limited to its 6 nearest ones.
The time scale $\tau$ governs the relaxation of the angular velocity $\varpi$ and thus the persistence of the perturbations felt by spin $i$. 
Note that with the above writing, $\varpi_i$ will become a white noise with thermal temperature $T$ in the equilibrium 
$\tau\to 0$ limit. At finite $\tau$, $T$ can be seen as
the local temperature of spinners and $\tau$ as some inverse angular friction coefficient.

\begin{figure*}[t!]
   \includegraphics[width=\textwidth]{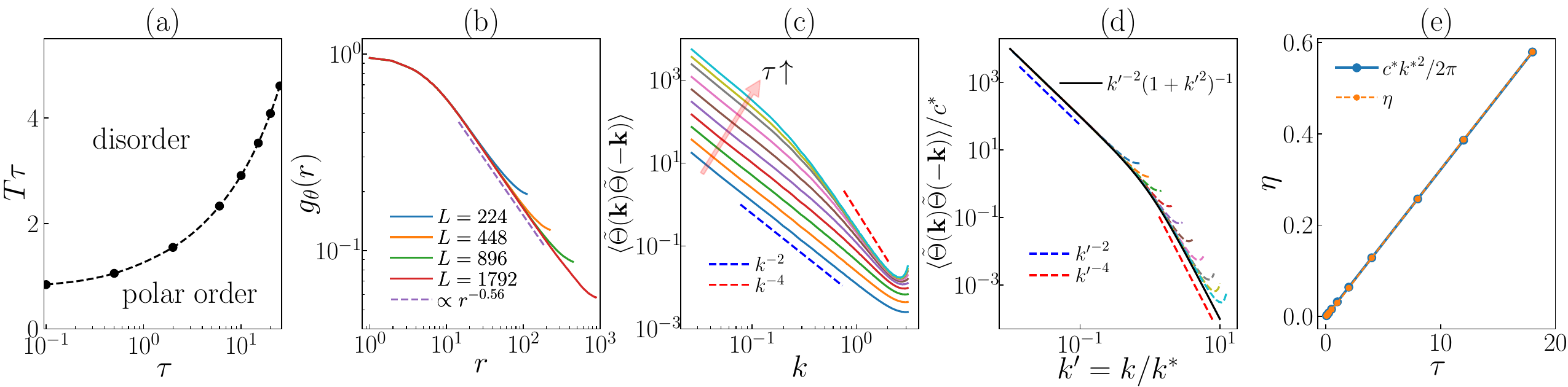}
    \caption{XY model with time-correlated noise (Eqs.~(\ref{XY},\ref{OU})), simulated on a triangular lattice
    containing $64\times64$ elementary blocks of $7\times 8$ sites making an almost-square 
    domain with about 230,000 sites. Periodic boundary conditions are used. Simple Euler explicit scheme, with timestep $0.01$, and typical simulation time of $10^7$.
    (a) phase diagram in the $(\tau, T)$ plane. Order-disorder transition defined to be at
   inflexion points of polar order parameter curves. 
    (b) angular correlation function $g_\theta(r)$ for $\tau=20$, $T=0.15$ and $\kappa=0.5$ and various system sizes.
    (c) spatial spectra of angular field at various $\tau$ values ($T=0.175$, $\kappa=0.5$). For curves from bottom to top, we have $\tau =0.06$, $0.12$, $0.25$, $0.5$, $1$, $2$, $4$, $8$, $12$, $18$.
    (d) same spectra as in (b), but rescaled as explained in the main text.
    (e) variation of decay exponent $\eta$ with $\tau$ from direct measurement as in (b), 
    and from expression $c^* {k^*}^2/2\pi$.
 }
\label{fig3}
\end{figure*}

This out-of-equilibrium model has been considered before in \cite{paoluzzi2018effective}, 
where the focus was mostly 
on the perturbative effects introduced in the small $\tau$ limit~\footnote{Note that the Ornstein-Uhlenbeck
process is written differently from our Eq.\eqref{OU} in \cite{paoluzzi2018effective}, so that $\tau$ is not
the same in both formulations. In \cite{paoluzzi2018effective}, the `large-scale temperature' is constant and small scales freeze as $\tau$ increases, whereas in our writing the `local temperature' is constant (for not too large $\tau$), and the large-scale one increases with $\tau$.
}.
Here we consider arbitrary values of $\tau$. A rough phase diagram, obtained at fixed system size, is given 
in Fig.~\ref{fig3}(a). As in equilibrium a quasi-long-range polar ordered phase with continuously-varying
exponents is observed at values of $T\tau$ which increase when $\tau$ is increased.

We first study the linear field theory that can be easily derived from Eqs.~(\ref{XY},\ref{OU}).
The orientation and angular rotation fields evolve according to:
\begin{align}
&\partial_t \Theta({\bf r},t) = \kappa_\theta \nabla^2\Theta({\bf r},t) + \Omega({\bf r},t) \label{Theta}\\
&\partial_t \Omega({\bf r},t) = -\frac{1}{\tau} \Omega({\bf r},t) +\sqrt{2\rho T/\tau} \, \Xi({\bf r},t) \label{Omega}
\end{align}
where the stiffness $\kappa_\theta = \sqrt{3} \kappa$ for a triangular lattice,
$\Xi({\bf r},t)$ is a Gaussian white field with $\langle \Xi({\bf r},t)\Xi({\bf r}',t')\rangle = 
\delta({\bf r}-{\bf r}') \delta(t-t')$, and $\rho$ is the lattice number density.
The angular rotation field $\Omega$ has a well defined kinetic energy that is independent of $\tau$ since 
$\langle |\Omega|^2\rangle = \rho T$.

The correlations of the angular field $\Theta$ can be obtained by 
solving Eqs.~(\ref{Theta},\ref{Omega}) in Fourier space. One finds in particular
\begin{equation}
\tilde{\Theta}({\bf k}, \omega) = \frac{\sqrt{2\rho T/\tau}}
{(i\omega +\kappa_\theta k^2)(i\omega+1/\tau)} \, \tilde{\Xi}({\bf k}, \omega)
\end{equation}
so that the autocorrelation 
\begin{equation}
\langle\tilde{\Theta}({\bf k}) \Tilde{\Theta}({\bf -k}) \rangle = 
\frac{\rho S/\kappa_\theta}{1+\kappa_\theta \tau k^2} \frac{\tau T}{k^2} \;,
\label{correl}
\end{equation}
where $S$ is the volume of the system.
Thus, on large scales, the effective temperature governing fluctuations is nothing but
$T_S = \tau T$. Integrating \eqref{correl} assuming a Gaussian distribution of angles yields the 
angular correlation function 
\begin{equation}
g_\theta (r)  = \langle \cos [ \Theta({\bf r}) - \Theta(0)] \rangle
 \underset{r\to\infty}{\sim} \left( \frac{r}{\ell_\tau}\right)^{-\eta} 
\end{equation} 
with
\begin{equation}
\eta\!=\!\frac{\tau T}{2\pi\kappa_\theta}
\label{eta}
\end{equation}
where $\ell_\tau \sim \sqrt{\kappa_\theta \tau}$ when it is much larger than the lattice spacing $\ell_0$.

We thus expect $\eta$ to increase linearly with either $T$ or $\tau$, while 
the spins kinetic temperature, defined as the angular rotational kinetic energy, is independent of $\tau$. 
Indeed $\tau$ drops from the large-$k$ limit of fluctuations:
\begin{equation}
\lim_{k\to\infty}\langle\tilde{\Theta}({\bf k}) \Tilde{\Theta}({\bf -k}) \rangle = 
\frac{\rho S T}{\kappa_\theta^2 k^4} \;.
\end{equation}
The associated `kinetic temperature' is expected to be responsible for the local events marking the order-disorder transition (such as the excitation and unbinding of topological defects).
Thus, at large $\tau$, we expect the large-scale temperature $T_S$ and the decay exponent $\eta$ 
to be able to reach large values at the transition.

The above results are confirmed numerically. 
First of all, $\eta$,
measured directly from the decay of the correlation function $g_\theta(r)$, 
is easily observed to take values much larger than $\tfrac{1}{4}$,
the value marking the Berezinskii-Kosterlitz-Thouless transition (Fig.~\ref{fig3}(b)). 

Spatial spectra of the angular field have the form predicted by Eq.~\eqref{correl}, developing a 
$\tau$-independent $1/k^4$ region at short scales while the amplitude of their large-scale $1/k^2$ 
region increases with $\tau$ (Fig.~\ref{fig3}(c)).
Using $k^*=1/\sqrt{\kappa_\theta \tau}$ and $c^*=\tau^2 T$, defining $k'=k/k^*$, the angular field spectra
obtained at different $\tau$ values collapse nearly perfectly onto the master curve $1/[k'^2 (1+k'^2)]$
(Fig.~\ref{fig3}(d)).

Direct estimates of $\eta$ values (from data such as Fig.~\ref{fig3}(b)) and $\eta$ values calculated from
 $c^* {k^*}^2/2\pi=\tau T/(2\pi\kappa_\theta)$ (the expression found in \eqref{eta}) coincide 
 and increase linearly with $\tau$ (Fig.~\ref{fig3}(e)).

The above results thus confirm the intuition gathered from our previous study of active crystals: 
the mere persistence in time of fluctuations extends the quasi-long-range orientational order region, 
allowing stronger spin waves without apparently triggering the nucleation and unbinding of defect pairs. 

{\it Passive crystal in an active bath.}
Coming back to 2D crystals, the case of the XY model subjected to time-correlated noise
suggests that extreme deformability
could be observed for passive crystals subjected to extrinsic persistent fluctuations. 
Think, say, of a colloidal crystal in a liquid containing small active particles or some microswimmers such as bacteria.

We now present a simple model of such a situation, where passive particles experiencing solely pairwise repulsion
between themselves are immersed in a bath of smaller active Brownian particles bumping into them. 
Passive particles are located initially at the nodes of a triangular lattice with lattice step $\ell_0=\tfrac{\sqrt{3}}{2}$ (density $\rho\simeq 1.54$), inside nearly-square domains accommodating an exact integer number of 
elementary cells, with periodic boundary conditions. 
For simplicity, the active particles, of number density $\rho_a$, do not interact between themselves, 
only with passive ones.

The positions ${\bf r}_i$ of the passive particles forming the crystal evolve according to
\begin{equation}
\dot{\bf r}_i =  \mu_r \sum_{j\sim i} (d_0-r_{ij}) \,{\bf e}_{ij} + \mu_a \sum_{a\sim i} (d_a-r_{ia}) \,{\bf e}_{ia} \;.
\label{bath1}
\end{equation}
The first term describes pairwise harmonic repulsion between particles of the crystal, with
$\mu_r$ their mobility, and where the sum is taken over particles $j$ 
within distance $d_0$ of $i$ ($d_0=1>\ell_0$),
${\bf e}_{ij}$ is the unit vector along ${\bf r}_j - {\bf r}_i$, and $r_{ij}=|{\bf r}_j - {\bf r}_i|$.
The second term stands for the repulsive forces --a priori of different nature-- exerted by active particles
(subscript ``a" stands for ``active") within distance $d_a=0.4<d_0$ of active particle $i$, 
with $\mu_a$ the mobility of passive particles under the
forces exerted on them by active particles, 
${\bf e}_{ia}$ the unit vector along the direction linking $i$ to the active particle $a$ 
and $r_{ia}$ the distance between them.
We consider a rather strong repulsion $\mu_r=10$ between passive particles, 
ensuring the integrity of the initial crystalline arrangement.

The positions ${\bf r}_a$ and polarities $\theta_a$ of the bath active particles follow:
\begin{align}
\dot{\bf r}_a &=  s_0 \,{\bf e}(\theta_a) + \mu_a \sum_{j\sim a} (d_a-r_{aj}) \,{\bf e}_{aj} \label{bath2}\\
\dot\theta_a &= \sqrt{2D_r}\,\xi_a(t) \label{bath3} \;,
\end{align}
where $D_r$ is the rotational diffusion strength, $\xi_i(t)$ is a zero-mean Gaussian white noise 
with $\langle\xi_i(t)\xi_j(t^\prime)\rangle=\delta_{ij}\delta(t-t^\prime)$, and the sum is taken
over all passive particles within distance $d_a$. 

\begin{figure}[t!]
    \includegraphics[width=\columnwidth]{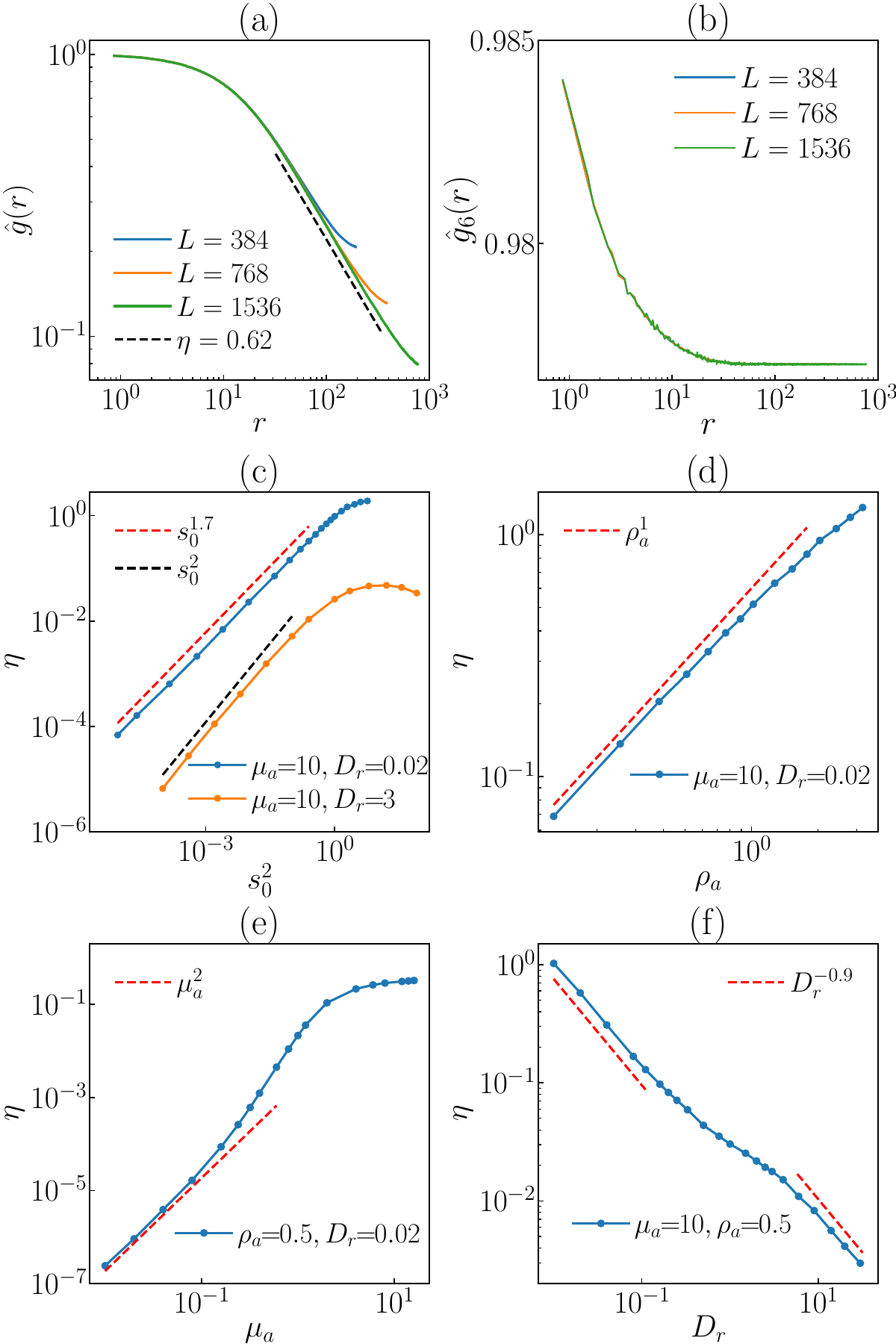}
    \caption{Passive crystal in an active bath (Eqs.(\ref{bath1}-\ref{bath3})). 
    Perfect initial triangular lattice made of $n\times n$ elementary blocks of $7\times 8$ sites making almost-square 
    domains. Periodic boundary conditions are used. Simple Euler explicit scheme, with timestep $0.01$, and typical simulation time of $10^7$.
        (a,b) positional and bond order correlation functions $\hat{g}(r)$ and $\hat{g}_6(r)$ 
        for various system sizes $L=8n$ ($\rho_a=0.637$, $\mu_a=10$, $D_r=0.02$, $s_0=0.7$).
        (c-f) variation with, respectively, $s_0^2$, $\rho_a$, $\mu_a$, and $D_r$ of the
        decay exponent $\eta$ of $\hat{g}(r)$ extracted from data such as in (b) for $L=384$ (parameters in legends, others as in (a)).}
\label{fig4}
\end{figure}

Below we only report on cases where the initial crystal does not melt and no defects arise. 
This allows to use, as in \cite{shi2023extreme}, the following non-oscillatory two-point correlation functions
for passive particles.
Positional order is estimated via 
\cite{engel2013hard}
\begin{equation}
\label{eq:g}
\hat{g}(r) = \left\langle \frac{\sum_{j\ne k}\delta(r-|\hat{\bf r}_j - \hat{\bf r}_k|) e^{i\hat{\bf G}\cdot[{\bf u}_j - {\bf u}_k]}}{\sum_{j\ne k}\delta(r-|\hat{\bf r}_j - \hat{\bf r}_k|)} \right\rangle_t
\end{equation}
where $\hat{\bf r}_i$ is the position of particle $i$ on the initial perfect lattice, 
${\bf u}_i(t)={\bf r}_i(t)-\hat{\bf r}_i$ its displacement vector,
and $\hat{\bf G}$ is one of the reciprocal vectors of the perfect lattice.

The correlation function $\hat{g}_6(r)$ of hexatic bond order
is defined similarly, replacing the exponential
by $\psi_6^*(j)\psi_6(k)$ with the local order 
$\psi_6(j)=\langle e^{i6\theta_{jj'}} \rangle_{j'\sim j}$ 
where $j'$ denotes the Voronoi neighbors of particle $j$ 
and $\theta_{jj'}$ is the orientation of ${\bf r}_{j'}-{\bf r}_j$.

One easily finds regimes where the initial crystal structure remains intact, no defects emerge, 
and yet $\eta$, the decay exponent of the two-point positional order correlation function, 
is larger than $\tfrac{1}{3}$ while bond order is long-range. 
Figure~\ref{fig4}(a,b) shows such an example, obtained with active particles
having relatively strong self-propulsion force $s_0$ and weak rotational diffusion $D_r$.
This confirms that a 2D crystal can exhibit strong deformations without melting 
as soon as its particles are subjected to persistent perturbations, be they intrinsic (crystal of active particles) or 
extrinsic (passive crystal in an active bath).

Next we explore further our passive crystal immersed in an active bath, providing insights
for future experimental realizations, and revealing properties heretofore not observed in crystals made of active 
particles.

As shown in Fig.~\ref{fig4}(c) (upper curve), the variation of $\eta$ with $s_0$ can deviate from the 
scaling $\eta \propto s_0^2$
observed in simple active crystals such as those studied in \cite{shi2023extreme}. 
In this case, for which the active particles are very persistent ($D_r=0.02$), we observe a slower growth law, reasonably well fitted, for small $s_0$ values, 
by a $\eta \propto s_0^{\alpha}$ powerlaw, but with $\alpha\simeq 1.7 < 2$.
Thus $s_0^2$ does not behave like an effective temperature in this case,
indicating that the external persistent perturbations exerted by the active particles can be
essentially different from the persistent noise permanently exerted by crystal particles. 
Note, however, that when $D_r$ is large, the $\eta \propto s_0^2$ scaling is observed
Fig.~\ref{fig4}(c) (lower curve).
That $\eta$ can grow slower than $s_0^2$ for low $D_r$ values can be rationalized as follows: 
increasing $s_0$ does not just increase the intensity of the external force
felt by crystal particles (as in the crystals made of active particles studied in \cite{shi2023extreme}), it also
shortens the typical period of time during which this force is felt, because of faster transverse sliding upon collision.
This explains intuitively why $\eta$ increases slower than $s_0^2$, but not, of course the $\sim1.7$
exponent observed, nor even whether the dependence of $\eta$ on $s_0$ is indeed algebraic. Other
cases (not shown) are best fitted by powerlaws with different $\alpha$ values. 
For large $D_r$, the `standard' scaling $\eta \propto s_0^2$ is observed because the interactions of the active bath particles with the passive crystal ones are short-lived and much more random. 
(Note that they are also weaker, hence the lower $\eta$ values observed.)
Finally, at very large $s_0$ values, $\eta$ grows slower and can even decrease because the active particles 
can `go through' passive ones under the effect of their strong self-propulsion force.

The density of active particles, $\rho_a$, has also, of course, a strong influence on the properties of the crystal
(Fig.~\ref{fig4}(d)). At small $\rho_a$ one observes, as expected (see, e.g. \eqref{correl}),
that $\eta \propto \rho_a$. As $\rho_a$ is increased, however, we observe a slower increase of $\eta$, 
probably due to the fact that then, typically, several active particles are in contact with a passive
crystal particle: their combined action results in a weaker force exerted.
(Note that in more realistic situations where the active particles do interact with each other, their combined action
could be further complicated, leading to different large-$\rho_a$ behavior.)

The mutual repulsive force strength $\mu_a$ between active and passive particles is also strongly influencing
the deformations exhibited by the crystal (Fig.~\ref{fig4}(e)). At small $\mu_a$, the bath particles can easily `penetrate' the crystal particles and escape from them, so that $\mu_a$ is very similar to
the intensity of the self-propelled force in simple active crystals. 
We thus expect (and observe) that $\eta \propto \mu_a^2$.
At large $\mu_a$ values, active particles can remain `stuck' to a passive particle for a while, 
increasing the effective 
persistence time of their action, and thus leading to a faster than $\mu_a^2$ increase of $\eta$. 
Finally, the value of the decay exponent $\eta$ saturates at large $\mu_a$ values, when the total effective force on crystal particles is completely dominated by the self-propulsion force of active particles.

We conclude by examining the dependence of the decay exponent $\eta$ on $D_r$, 
the rotational diffusion constant of the active bath particles. For the simple active crystals studied in
\cite{shi2023extreme}, we observed that $\eta\propto 1/D_r$, something not too surprising since $1/D_r$ 
is akin the persistence time $\tau$ considered in the first part of the present work.
For our passive crystal immersed in an active bath, we observe a slightly slower algebraic decay, 
with an exponent $\sim 0.9$, possibly for reasons similar to those giving the $s_0^{1.7}$ in panel (c)
(Fig.~\ref{fig4}(f)).
Importantly, this regime is `disturbed' at intermediate values $D_r\in [\sim0.5; \sim4]$, 
corresponding to lengthscales $s_0/D_r$ of the order of typical elementary scales in our system, 
where a weaker variation is observed. Note that even in the large-$D_r$ regime, where we expect our system to 
reach equilibrium, we do not recover the simple $1/D_r$ behavior.

{\it Discussion.}
The results above show that time-persistent external perturbations can have a profound influence
on the quasi-long-range ordered phases exhibited by 2D systems, 
as hinted recently in our study of simple active crystals \cite{shi2023extreme}.

In the simplest case of the XY model subjected to Ornstein-Uhlenbeck noise, strong spinwaves leading to fast
decay of orientational order can be observed without the system apparently entering the disordered phase. 
Whether the order-disorder transition remains of the Berezinskii-Kosterlitz-Thouless type is a key question left 
for future work, alongside that of the melting transition of 2D active crystals.

We studied next a 2D passive crystal immersed in a bath of active particles, 
a situation we believe to be of 
experimental relevance. We showed there also that the persistent forces exerted by the active bath particles 
on the crystal can induce large deformations without melting it. Our results reveal, however, that the theoretical
picture developed in \cite{shi2023extreme} for simple crystals made of active particles needs to be refined
to account for the effects of the active bath.
In particular we found significant deviations from the simple scaling
$\eta \sim s_0^2/D_r$ found in \cite{shi2023extreme} (Fig.~\ref{fig4}(c,f)).
We interpret these deviations as evidence that the effective noise exerted on a 2D crystal by an active bath is more complex, especially since the crystal can itself influence the surrounding bath. 
Our results call for an appropriate theoretical treatment, at continuum level,
of the coupling between the bath and the crystal, possibly inspired by the results of 
\cite{baek2018generic,granek2020bodies,granek2022anomalous,solon2022einstein}.

\acknowledgments
We thank Yu Duan, Beno{\^{\i}}t Mahault, Alexandre Solon, and Yongfeng Zhao for useful remarks and a careful reading of the manuscript.
This work is supported by the National Natural Science Foundation of China (Grants No. 12275188 and No. 11922506)

\bibliographystyle{apsrev4-2}
\bibliography{./Biblio-current.bib}

\end{document}